\input harvmac
\input epsf
%\draftmode

%%%%%%%%%%%%%%%%%%%%%%%%%%%%%%%%%%%%%%%%%%%%%%%%%%%%%%%%%%%%%%%%%%%%%%%%
%%%%%%%%%%%%%%%%%%%%%%%%%%%%%%%%%%%%%%%%%%%%%%%%%%%%%%%%%%%%%%%%%%%%%%%%
%	Emil Martinec's macros
%
%

\noblackbox
%

% Something to deal with sub-sub-sections

\def\unlockat{\catcode`\@=11}
\def\lockat{\catcode`\@=12}

\unlockat
% Something to deal with sub-sub-sections

\def\newsec#1{\global\advance\secno by1\message{(\the\secno. #1)}
\global\subsecno=0\global\subsubsecno=0\eqnres@t\noindent
{\bf\the\secno. #1}
\writetoca{{\secsym} {#1}}\par\nobreak\medskip\nobreak}
\global\newcount\subsecno \global\subsecno=0
\def\subsec#1{\global\advance\subsecno
by1\message{(\secsym\the\subsecno. #1)}
\ifnum\lastpenalty>9000\else\bigbreak\fi\global\subsubsecno=0
\noindent{\it\secsym\the\subsecno. #1}
\writetoca{\string\quad {\secsym\the\subsecno.} {#1}}
\par\nobreak\medskip\nobreak}
\global\newcount\subsubsecno \global\subsubsecno=0
\def\subsubsec#1{\global\advance\subsubsecno by1
\message{(\secsym\the\subsecno.\the\subsubsecno. #1)}
\ifnum\lastpenalty>9000\else\bigbreak\fi
\noindent\quad{\secsym\the\subsecno.\the\subsubsecno.}{#1}
\writetoca{\string\qquad{\secsym\the\subsecno.\the\subsubsecno.}{#1}}
\par\nobreak\medskip\nobreak}

\def\subsubseclab#1{\DefWarn#1\xdef
#1{\noexpand\hyperref{}{subsubsection}%
{\secsym\the\subsecno.\the\subsubsecno}%
{\secsym\the\subsecno.\the\subsubsecno}}%
\writedef{#1\leftbracket#1}\wrlabeL{#1=#1}}% Macros for boxes
\lockat

%

%% MORE MACROS

\font\manual=manfnt \def\dbend{\lower3.5pt\hbox{\manual\char127}}

\def\IZ{\relax\ifmmode\mathchoice
{\hbox{\cmss Z\kern-.4em Z}}{\hbox{\cmss Z\kern-.4em Z}}
{\lower.9pt\hbox{\cmsss Z\kern-.4em Z}}
{\lower1.2pt\hbox{\cmsss Z\kern-.4em Z}}\else{\cmss Z\kern-.4em
Z}\fi}

% more macros, alphabetically

\def\IZ{\relax\ifmmode\mathchoice
{\hbox{\cmss Z\kern-.4em Z}}{\hbox{\cmss Z\kern-.4em Z}}
{\lower.9pt\hbox{\cmsss Z\kern-.4em Z}}
{\lower1.2pt\hbox{\cmsss Z\kern-.4em Z}}\else{\cmss Z\kern-.4em
Z}\fi}
\def\IB{\relax{\rm I\kern-.18em B}}
\def\IC{{\relax\hbox{$\inbar\kern-.3em{\rm C}$}}}
\def\ID{\relax{\rm I\kern-.18em D}}
\def\IE{\relax{\rm I\kern-.18em E}}
\def\IF{\relax{\rm I\kern-.18em F}}
\def\IG{\relax\hbox{$\inbar\kern-.3em{\rm G}$}}
\def\IGa{\relax\hbox{${\rm I}\kern-.18em\Gamma$}}
\def\IH{\relax{\rm I\kern-.18em H}}
\def\II{\relax{\rm I\kern-.18em I}}
\def\IK{\relax{\rm I\kern-.18em K}}
\def\IP{\relax{\rm I\kern-.18em P}}
\def\IQ{\relax\hbox{$\inbar\kern-.3em{\rm Q}$}}

\def\IS{{\bf S}}
\def\IT{{\bf T}}

\def\inbar{\,\vrule height1.5ex width.4pt depth0pt}

\font\cmss=cmss10 \font\cmsss=cmss10 at 7pt
\def\IR{\relax{\rm I\kern-.18em R}}

% Macros for boxes
%
%\def\boxit#1{\vbox{\hrule\hbox{\vrule\kern8pt
%\vbox{\hbox{\kern8pt}\hbox{\vbox{#1}}\hbox{\k
%\hbox{$\displaystyle #1$}\kern8pt}\kern8pt\vrule}\hrule}}}
%
%%% MACROS FOR BOX BOUNDARY CONDS
%%% FROM KAWAI ET AL

%\def\makeblankbox#1#2{\hbox{\lower\dp0\vbox{\hidehrule{#1}{#2}%
%   \kern -#1% overlap rules
%   \hbox to \wd0{\hidevrule{#1}{#2}%
%      \raise\ht0\vbox to #1{}% vrule height
%      \lower\dp0\vtop to #1{}% vrule depth
%      \hfil\hidevrule{#2}{#1}}%
%   \kern-#1\hidehrule{#2}{#1}}}%
%}%
%\def\hidehrule#1#2{\kern-#1\hrule height#1 depth#2 \kern-#2}%
%\def\hidevrule#1#2{\kern-#1{\dimen0=#1\advance\dimen0 by #2\vrule
%    width\dimen0}\kern-#2}%
%\def\openbox{\ht0=1.2mm \dp0=1.2mm \wd0=2.4mm  \raise 2.75pt
%\makeblankbox {.25pt} {.25pt}  }
%\def\qed{\hskip 8mm \openbox}
%\def\abs#1{\left\vert #1 \right\vert}
%\def\bun#1/#2{\leavevmode
%   \kern.1em \raise .5ex \hbox{\the\scriptfont0 #1}%
%   \kern-.1em $/$%
%   \kern-.15em \lower .25ex \hbox{\the\scriptfont0 #2}%
%}
%\def\row#1#2{#1_1,\ldots,#1_#2}
%\def\apar{\noalign{\vskip 2mm}}
%\def\blackbox{\hbox{\vrule height .5ex width .3ex depth -.3ex}}
%\def\nord{{\textstyle {\blackbox\atop\blackbox}}}
%\def\ts{\,}
%\def\opensquare{\ht0=3.4mm \dp0=3.4mm \wd0=6.8mm  \raise 2.7pt
%\makeblankbox {.25pt} {.25pt}  }

%%%%%%%%%%%%%%%%%%%%%%%

\def\sector#1#2{\ {\scriptstyle #1}\hskip 1mm
\mathop{\opensquare}\limits_{\lower 1mm\hbox{$\scriptstyle#2$}}\hskip 1mm}

\def\tsector#1#2{\ {\scriptstyle #1}\hskip 1mm
\mathop{\opensquare}\limits_{\lower 1mm\hbox{$\scriptstyle#2$}}^\sim\hskip 1mm}
%%%
%%%

%% ANOTHER SET OF MACROS

\def\inbar{\,\vrule height1.5ex width.4pt depth0pt}

\font\cmss=cmss10 \font\cmsss=cmss10 at 7pt
\def\IR{\relax{\rm I\kern-.18em R}}

%% new macros

\def\frac#1#2{{#1\over#2}}

\def\inbar{\,\vrule height1.5ex width.4pt depth0pt}
\def\IC{\relax\hbox{$\inbar\kern-.3em{\rm C}$}}
\def\IR{\relax{\rm I\kern-.18em R}}
\def\IP{\relax{\rm I\kern-.18em P}}

%
%%%%%%%%%%%%%%%%%%%%%%%%%%%%%%%%%%%%
%
\catcode`\@=11
\def\slash#1{\mathord{\mathpalette\c@ncel{#1}}}
\overfullrule=0pt

\def\II{{\cal I}}

\def\NN{{\cal N}}

\def\underrel#1\over#2{\mathrel{\mathop{\kern\z@#1}\limits_{#2}}}

\catcode`\@=12

%%%%%%%%%%%%%%%%%%%%%%%%%%%%%%%%%%%%%%%%%%%%%%%%%%%%%%%%%%%%%%

%

\def\exp{{\rm exp}}

%%%%%%%%%%%%%%%%%%%%%%%%%%%%%%%%%%%%%%%%%%%%%%%%%%%%%%%%%%%%%%
% new defs:

\def\ra{{\rightarrow}}

\def\F{{\cal F}}
\def\X{{\bf X}}

%
%%%%%%%%%%%%%%%%%%%%%%%%%%%%%%%%%%%%%%%%%%%%%%%%%%%%%%%%%%%%%%%%%%%%%%%%%%

%% END MACROS
%%

%%%%%%%%%%%%%%%%%%%%%%%%%%%%%%%%%%%%%%%%%%%%%%%%%%%%%%%%%%%%%%%%%%%%%%%%%%
%%%%%%%%%%%%%%%%%%%%%%%%%%%%%%%%%%%%%%%%%%%%%%%%%%%%%%%%%%%%%%%%%%%%%%%%%%
%%%
%%% References
%%%
%%%%%%%%%%%%%%%%%%%%%%%%%%%%%%%%%%%%%%%%%%%%%%%%%%%%%%%%%%%%%%%%%%%%%%%%%%
%%%%%%%%%%%%%%%%%%%%%%%%%%%%%%%%%%%%%%%%%%%%%%%%%%%%%%%%%%%%%%%%%%%%%%%%%%

%\KutasovJP
\lref\KutasovJP{
  D.~Kutasov and D.~A.~Sahakyan,
  ``Comments on the thermodynamics of little string theory,''
  JHEP {\bf 0102}, 021 (2001)
  [arXiv:hep-th/0012258].
  %%CITATION = HEP-TH 0012258;%%
}

%\KutasovRR
\lref\KutasovRR{
  D.~Kutasov,
  ``Accelerating branes and the string / black hole transition,''
  arXiv:hep-th/0509170.
  %%CITATION = HEP-TH 0509170;%%
}

%\GiveonMI
\lref\GiveonMI{
  A.~Giveon, D.~Kutasov, E.~Rabinovici and A.~Sever,
  ``Phases of quantum gravity in AdS(3) and linear dilaton backgrounds,''
  Nucl.\ Phys.\ B {\bf 719}, 3 (2005)
  [arXiv:hep-th/0503121].
  %%CITATION = HEP-TH 0503121;%%
}

%\GiveonJV
\lref\GiveonJV{
  A.~Giveon and D.~Kutasov,
  ``The charged black hole / string transition,''
  arXiv:hep-th/0510211.
  %%CITATION = HEP-TH 0510211;%%
}

%\MizoguchiKK
\lref\MizoguchiKK{
  S.~Mizoguchi,
  ``Modular invariant critical superstrings on four-dimensional Minkowski
  space x two-dimensional black hole,''
  JHEP {\bf 0004}, 014 (2000)
  [arXiv:hep-th/0003053].
  %%CITATION = HEP-TH 0003053;%%
}

%\KutasovUA
\lref\KutasovUA{
  D.~Kutasov and N.~Seiberg,
  ``Noncritical Superstrings,''
  Phys.\ Lett.\ B {\bf 251}, 67 (1990).
  %%CITATION = PHLTA,B251,67;%%
}

%\ItzhakiZR
\lref\ItzhakiZR{
  N.~Itzhaki, D.~Kutasov and N.~Seiberg,
  ``Non-supersymmetric deformations of non-critical superstrings,''
  arXiv:hep-th/0510087.
  %%CITATION = HEP-TH 0510087;%%
}

%\SusskindAA
\lref\SusskindAA{
  L.~Susskind,
  %``Strings, black holes and Lorentz contraction,''
  Phys.\ Rev.\ D {\bf 49}, 6606 (1994)
  [arXiv:hep-th/9308139]; 
%\SusskindWS
%\lref\SusskindWS{
  %L.~Susskind,
  ``Some speculations about black hole entropy in string theory,''
  arXiv:hep-th/9309145.
  %%CITATION = HEP-TH 9309145;%%
  %%CITATION = HEP-TH 9308139;%%
}

%\HorowitzNW
\lref\HorowitzNW{
  G.~T.~Horowitz and J.~Polchinski,
  %``A correspondence principle for black holes and strings,''
  Phys.\ Rev.\ D {\bf 55}, 6189 (1997)
  [arXiv:hep-th/9612146];
  %%CITATION = HEP-TH 9612146;%%
%\HorowitzJC
%\lref\HorowitzJC{
%  G.~T.~Horowitz and J.~Polchinski,
  ``Self gravitating fundamental strings,''
  Phys.\ Rev.\ D {\bf 57}, 2557 (1998)
  [arXiv:hep-th/9707170].
  %%CITATION = HEP-TH 9707170;%%
}

%\AtickSI
\lref\AtickSI{
  J.~J.~Atick and E.~Witten,
  ``The Hagedorn Transition And The Number Of Degrees Of Freedom Of String
  Theory,''
  Nucl.\ Phys.\ B {\bf 310}, 291 (1988).
  %%CITATION = NUPHA,B310,291;%%
}

%\DienesHX
\lref\DienesHX{
  K.~R.~Dienes, E.~Dudas, T.~Gherghetta and A.~Riotto,
  ``Cosmological phase transitions and radius stabilization in higher
  dimensions,''
  Nucl.\ Phys.\ B {\bf 543}, 387 (1999)
  [arXiv:hep-ph/9809406].
  %%CITATION = HEP-PH 9809406;%%
}

%\GreenSP
\lref\GreenSP{
  M.~B.~Green, J.~H.~Schwarz and E.~Witten,
  ``Superstring Theory. Vol. 1: Introduction,''
%\href{http://www.slac.stanford.edu/spires/find/hep/www?irn=1755021}{SPIRES entry}
}

%\PolchinskiZF
\lref\PolchinskiZF{
  J.~Polchinski,
  ``Evaluation Of The One Loop String Path Integral,''
  Commun.\ Math.\ Phys.\  {\bf 104}, 37 (1986).
  %%CITATION = CMPHA,104,37;%%
}

%\OoguriZV
\lref\OoguriZV{
  H.~Ooguri, A.~Strominger and C.~Vafa,
  ``Black hole attractors and the topological string,''
  Phys.\ Rev.\ D {\bf 70}, 106007 (2004)
  [arXiv:hep-th/0405146].
  %%CITATION = HEP-TH 0405146;%%
}

%more refs

%\GiveonRW
\lref\GiveonRW{
  A.~Giveon, A.~Konechny, E.~Rabinovici and A.~Sever,
  ``On thermodynamical properties of some coset CFT backgrounds,''
  JHEP {\bf 0407}, 076 (2004)
  [arXiv:hep-th/0406131].
  %%CITATION = HEP-TH 0406131;%%
}

%\KutasovJP
\lref\KS{
  D.~Kutasov and D.~A.~Sahakyan,
  ``Comments on the thermodynamics of little string theory,''
  JHEP {\bf 0102}, 021 (2001)
  [arXiv:hep-th/0012258].
  %%CITATION = HEP-TH 0012258;%%
}

%\GiveonTQ
\lref\GKb{
  A.~Giveon and D.~Kutasov,
  ``Comments on double scaled little string theory,''
  JHEP {\bf 0001}, 023 (2000)
  [arXiv:hep-th/9911039].
  %%CITATION = HEP-TH 9911039;%%
}

%\GiveonPX
\lref\GKa{
  A.~Giveon and D.~Kutasov,
  ``Little string theory in a double scaling limit,''
  JHEP {\bf 9910}, 034 (1999)
  [arXiv:hep-th/9909110].
  %%CITATION = HEP-TH 9909110;%%
}

%\AharonyVK
\lref\AFKS{
  O.~Aharony, B.~Fiol, D.~Kutasov and D.~A.~Sahakyan,
  ``Little string theory and heterotic/type II duality,''
  Nucl.\ Phys.\ B {\bf 679}, 3 (2004)
  [arXiv:hep-th/0310197].
  %%CITATION = HEP-TH 0310197;%%
}

%\GiveonWN
\lref\GKPS{
  A.~Giveon, A.~Konechny, A.~Pakman and A.~Sever,
  ``Type 0 strings in a 2-d black hole,''
  JHEP {\bf 0310}, 025 (2003)
  [arXiv:hep-th/0309056].
  %%CITATION = HEP-TH 0309056;%%
}

%\NarayanDR
\lref\NarayanDR{
  K.~Narayan and M.~Rangamani,
  ``Hot little string correlators: A view from supergravity,''
  JHEP {\bf 0108}, 054 (2001)
  [arXiv:hep-th/0107111].
  %%CITATION = HEP-TH 0107111;%%
}

%\DeBoerDD
\lref\DeBoerDD{
  P.~A.~DeBoer and M.~Rozali,
  ``Thermal correlators in little string theory,''
  Phys.\ Rev.\ D {\bf 67}, 086009 (2003)
  [arXiv:hep-th/0301059].
  %%CITATION = HEP-TH 0301059;%%
}

%\HorowitzCD
\lref\HorowitzCD{
  G.~T.~Horowitz and A.~Strominger,
  ``Black strings and P-branes,''
  Nucl.\ Phys.\ B {\bf 360}, 197 (1991).
  %%CITATION = NUPHA,B360,197;%%
}

%\MaldacenaCG
\lref\MaldacenaCG{
  J.~M.~Maldacena and A.~Strominger,
  ``Semiclassical decay of near-extremal fivebranes,''
  JHEP {\bf 9712}, 008 (1997)
  [arXiv:hep-th/9710014].
  %%CITATION = HEP-TH 9710014;%%
}

%\SonSD
\lref\SonSD{
  D.~T.~Son and A.~O.~Starinets,
  ``Minkowski-space correlators in AdS/CFT correspondence: Recipe and
  applications,''
  JHEP {\bf 0209}, 042 (2002)
  [arXiv:hep-th/0205051].
  %%CITATION = HEP-TH 0205051;%%
}

%\PolicastroSE
\lref\PSS{
  G.~Policastro, D.~T.~Son and A.~O.~Starinets,
  ``From AdS/CFT correspondence to hydrodynamics,''
  JHEP {\bf 0209}, 043 (2002)
  [arXiv:hep-th/0205052].
  %%CITATION = HEP-TH 0205052;%%
}

%\KovtunEV
\lref\KST{
  P.~K.~Kovtun and A.~O.~Starinets,
  ``Quasinormal modes and holography,''
  arXiv:hep-th/0506184.
  %%CITATION = HEP-TH 0506184;%%
}

%
%
%\lref\KST{P.~K.~Kovtun and A.~O.~Starinets, "Quasinormal modes and
%holography", to appear.}
%

%\HarmarkHW
\lref\HarmarkHW{
  T.~Harmark and N.~A.~Obers,
  ``Hagedorn behaviour of little string theory from string corrections to
  NS5-branes,''
  Phys.\ Lett.\ B {\bf 485}, 285 (2000)
  [arXiv:hep-th/0005021].
  %%CITATION = HEP-TH 0005021;%%
}

%\BerkoozMZ
\lref\BerkoozMZ{
  M.~Berkooz and M.~Rozali,
  ``Near Hagedorn dynamics of NS fivebranes, or a new universality class  of
  coiled strings,''
  JHEP {\bf 0005}, 040 (2000)
  [arXiv:hep-th/0005047].
  %%CITATION = HEP-TH 0005047;%%
}

%\KovtunWP
\lref\KSSa{
  P.~Kovtun, D.~T.~Son and A.~O.~Starinets,
  ``Holography and hydrodynamics: Diffusion on stretched horizons,''
  JHEP {\bf 0310}, 064 (2003)
  [arXiv:hep-th/0309213].
  %%CITATION = HEP-TH 0309213;%%
}

%\KovtunDE
\lref\KSSb{
  P.~Kovtun, D.~T.~Son and A.~O.~Starinets,
  ``Viscosity in strongly interacting quantum field theories from black hole
  physics,''
  Phys.\ Rev.\ Lett.\  {\bf 94}, 111601 (2005)
  [arXiv:hep-th/0405231].
  %%CITATION = HEP-TH 0405231;%%
}

%\StarinetsBR
\lref\StarinetsBR{
  A.~O.~Starinets,
  ``Quasinormal modes of near extremal black branes,''
  Phys.\ Rev.\ D {\bf 66}, 124013 (2002)
  [arXiv:hep-th/0207133].
  %%CITATION = HEP-TH 0207133;%%
}

%\AharonyTT
\lref\AharonyTT{
  O.~Aharony and T.~Banks,
  ``Note on the quantum mechanics of M theory,''
  JHEP {\bf 9903}, 016 (1999)
  [arXiv:hep-th/9812237].
  %%CITATION = HEP-TH 9812237;%%
}

%\BerkoozMZ
\lref\BerkoozMZ{
  M.~Berkooz and M.~Rozali,
  ``Near Hagedorn dynamics of NS fivebranes, or a new universality class  of
  coiled strings,''
  JHEP {\bf 0005}, 040 (2000)
  [arXiv:hep-th/0005047].
  %%CITATION = HEP-TH 0005047;%%
}

%\RangamaniIR
\lref\RangamaniIR{
  M.~Rangamani,
  ``Little string thermodynamics,''
  JHEP {\bf 0106}, 042 (2001)
  [arXiv:hep-th/0104125].
  %%CITATION = HEP-TH 0104125;%%
}

%\BuchelDG
\lref\BuchelDG{
  A.~Buchel,
  ``On the thermodynamic instability of LST,''
  arXiv:hep-th/0107102.
  %%CITATION = HEP-TH 0107102;%%
}

%\SeibergZK
\lref\Seiberg{
  N.~Seiberg,
  ``New theories in six dimensions and matrix description of M-theory on  T**5
  and T**5/Z(2),''
  Phys.\ Lett.\ B {\bf 408}, 98 (1997)
  [arXiv:hep-th/9705221].
  %%CITATION = HEP-TH 9705221;%%
}

%\BenincasaIV
\lref\bbs{
  P.~Benincasa, A.~Buchel and A.~O.~Starinets,
  ``Sound waves in strongly coupled non-conformal gauge theory plasma,''
  arXiv:hep-th/0507026.
  %%CITATION = HEP-TH 0507026;%%
}

%
%\lref\bbs{ P.~Benincasa, A.~Buchel and A.O.~Starinets, "Sound
%waves in strongly coupled non-conformal gauge theory plasma", to
%appear. }
%

%\BuchelTZ
\lref\BuchelTZ{
  A.~Buchel and J.~T.~Liu,
  ``Universality of the shear viscosity in supergravity,''
  Phys.\ Rev.\ Lett.\  {\bf 93}, 090602 (2004)
  [arXiv:hep-th/0311175].
  %%CITATION = HEP-TH 0311175;%%
}

%\BuchelQQ
\lref\BuchelQQ{
  A.~Buchel,
  ``On universality of stress-energy tensor correlation functions in
  %supergravity,''
  Phys.\ Lett.\ B {\bf 609}, 392 (2005)
  [arXiv:hep-th/0408095].
  %%CITATION = HEP-TH 0408095;%%
}

%\HerzogFN
\lref\HerzogFN{
  C.~P.~Herzog,
  ``The hydrodynamics of M-theory,''
  JHEP {\bf 0212}, 026 (2002)
  [arXiv:hep-th/0210126].
  %%CITATION = HEP-TH 0210126;%%
}

%\AharonyTT
\lref\AharonyTT{
  O.~Aharony and T.~Banks,
  ``Note on the quantum mechanics of M theory,''
  JHEP {\bf 9903}, 016 (1999)
  [arXiv:hep-th/9812237].
  %%CITATION = HEP-TH 9812237;%%
}

%\RangamaniIR
\lref\RangamaniIR{
  M.~Rangamani,
  ``Little string thermodynamics,''
  JHEP {\bf 0106}, 042 (2001)
  [arXiv:hep-th/0104125].
  %%CITATION = HEP-TH 0104125;%%
}

%\BuchelDG
\lref\BuchelDG{
  A.~Buchel,
  ``On the thermodynamic instability of LST,''
  arXiv:hep-th/0107102.
  %%CITATION = HEP-TH 0107102;%%
}

%\AharonyXN
\lref\AharonyXN{
  O.~Aharony, A.~Giveon and D.~Kutasov,
  ``LSZ in LST,''
  Nucl.\ Phys.\ B {\bf 691}, 3 (2004)
  [arXiv:hep-th/0404016].
  %%CITATION = HEP-TH 0404016;%%
}

%\GiveonZM
\lref\GiveonZM{
  A.~Giveon, D.~Kutasov and O.~Pelc,
  ``Holography for non-critical superstrings,''
  JHEP {\bf 9910}, 035 (1999)
  [arXiv:hep-th/9907178].
  %%CITATION = HEP-TH 9907178;%%
}

%\OoguriWJ
\lref\OoguriWJ{
  H.~Ooguri and C.~Vafa,
  ``Two-Dimensional Black Hole and Singularities of CY Manifolds,''
  Nucl.\ Phys.\ B {\bf 463}, 55 (1996)
  [arXiv:hep-th/9511164].
  %%CITATION = HEP-TH 9511164;%%
}

%\KutasovTE
\lref\KutasovTE{
  D.~Kutasov,
  ``Orbifolds and Solitons,''
  Phys.\ Lett.\ B {\bf 383}, 48 (1996)
  [arXiv:hep-th/9512145].
  %%CITATION = HEP-TH 9512145;%%
}

%\AharonyUB
\lref\AharonyUB{
  O.~Aharony, M.~Berkooz, D.~Kutasov and N.~Seiberg,
  ``Linear dilatons, NS5-branes and holography,''
  JHEP {\bf 9810}, 004 (1998)
  [arXiv:hep-th/9808149].
  %%CITATION = HEP-TH 9808149;%%
}

%\MinwallaXI
\lref\MinwallaXI{
  S.~Minwalla and N.~Seiberg,
  ``Comments on the IIA NS5-brane,''
  JHEP {\bf 9906}, 007 (1999)
  [arXiv:hep-th/9904142].
  %%CITATION = HEP-TH 9904142;%%
}

%\PeetWN
\lref\PeetWN{
  A.~W.~Peet and J.~Polchinski,
  ``UV/IR relations in AdS dynamics,''
  Phys.\ Rev.\ D {\bf 59}, 065011 (1999)
  [arXiv:hep-th/9809022].
  %%CITATION = HEP-TH 9809022;%%
}

%\KapustinCI
\lref\KapustinCI{
  A.~Kapustin,
  ``On the universality class of little string theories,''
  Phys.\ Rev.\ D {\bf 63}, 086005 (2001)
  [arXiv:hep-th/9912044].
  %%CITATION = HEP-TH 9912044;%%
}

%\AtickSI
\lref\AtickSI{
  J.~J.~Atick and E.~Witten,
  ``The Hagedorn Transition And The Number Of Degrees Of Freedom Of String
  Theory,''
  Nucl.\ Phys.\ B {\bf 310}, 291 (1988).
  %%CITATION = NUPHA,B310,291;%%
}

%\BuchelDI
\lref\BuchelDI{
  A.~Buchel, J.~T.~Liu and A.~O.~Starinets,
  ``Coupling constant dependence of the shear viscosity in N = 4 supersymmetric
  Yang-Mills theory,''
  Nucl.\ Phys.\ B {\bf 707}, 56 (2005)
  [arXiv:hep-th/0406264].
  %%CITATION = HEP-TH 0406264;%%
}

%\KlemmBJ
\lref\KlemmBJ{
  A.~Klemm, W.~Lerche, P.~Mayr, C.~Vafa and N.~P.~Warner,
  ``Self-Dual Strings and N=2 Supersymmetric Field Theory,''
  Nucl.\ Phys.\ B {\bf 477}, 746 (1996)
  [arXiv:hep-th/9604034].
  %%CITATION = HEP-TH 9604034;%%
}

%\MaldacenaHW
\lref\MaldacenaHW{
  J.~M.~Maldacena and H.~Ooguri,
  ``Strings in AdS(3) and SL(2,R) WZW model. I,''
  J.\ Math.\ Phys.\  {\bf 42}, 2929 (2001)
  [arXiv:hep-th/0001053].
  %%CITATION = HEP-TH 0001053;%%
}

%\ArgyresXN
\lref\ArgyresXN{
  P.~C.~Argyres, M.~Ronen Plesser, N.~Seiberg and E.~Witten,
  ``New N=2 Superconformal Field Theories in Four Dimensions,''
  Nucl.\ Phys.\ B {\bf 461}, 71 (1996)
  [arXiv:hep-th/9511154].
  %%CITATION = HEP-TH 9511154;%%
}

%\ArgyresJJ
\lref\ArgyresJJ{
  P.~C.~Argyres and M.~R.~Douglas,
  ``New phenomena in SU(3) supersymmetric gauge theory,''
  Nucl.\ Phys.\ B {\bf 448}, 93 (1995)
  [arXiv:hep-th/9505062].
  %%CITATION = HEP-TH 9505062;%%
}

%\EguchiVU
\lref\EguchiVU{
  T.~Eguchi, K.~Hori, K.~Ito and S.~K.~Yang,
  ``Study of $N=2$ Superconformal Field Theories in $4$ Dimensions,''
  Nucl.\ Phys.\ B {\bf 471}, 430 (1996)
  [arXiv:hep-th/9603002].
  %%CITATION = HEP-TH 9603002;%%
}

%\DamourAW
\lref\DamourAW{
  T.~Damour and G.~Veneziano,
  ``Self-gravitating fundamental strings and black holes,''
  Nucl.\ Phys.\ B {\bf 568}, 93 (2000)
  [arXiv:hep-th/9907030].
  %%CITATION = HEP-TH 9907030;%%
}

%%%%%%%%%%%%%%%%%%%%%%%%%%%%%%%%%%%%%%%%%%%%%%%%%%%%%%%%%%%%%%
% paper starts here !!!

\Title{\vbox{\baselineskip12pt
\hbox{hep-th/0512075}
\hbox{NSF-KITP-05-111}
}}
{\vbox{\centerline{On Non-Critical Superstring/Black Hole Transition}
}}
\centerline{Andrei Parnachev and David A. Sahakyan}
\bigskip
\centerline{{\it Department of Physics, Rutgers University}}
\centerline{\it Piscataway, NJ 08854-8019, USA}
\centerline{and} 
\centerline{{\it Kavli Institute for Theoretical Physics, University of California}}
\centerline{\it Santa Barbara, CA 93106, USA}
\vskip.1in \vskip.1in \centerline{\bf Abstract}  
\noindent
An interesting case of string/black hole transition  occurs
in two-dimensional non-critical string theory dressed with
a compact CFT.
In these models the high energy densities of states of perturbative
strings and black holes have the same leading behavior when
the Hawking temperature of the black hole is equal to the
Hagedorn temperature of perturbative strings.
We compare the first subleading terms in the black hole and
closed string entropies in this setting and argue that the entropy interpolates
between these expressions  as the energy is varied.
We compute the subleading correction to the black hole
entropy for a specific simple model.

\vfill

\Date{Dec. 6, 2005}
   
%\draftmode

\newsec{Introduction and Summary}
The microscopic origin of black hole entropy is a fundamental
question to which string theory provides many clues.
For technical reasons, it is often easier to make
precise statements about BPS black holes at zero temperature.
Their  finite-temperature cousins, although harder to study,
are nonetheless  interesting in their own right.
Consider the Schwarzschild black hole in four dimensions, whose
Schwarzschild radius is proportional to the mass $r\sim M l_p^2$
(where $l_p$ is the Planck length).
When the mass (energy) is very large, the string corrections to
classical gravity are negligible, the entropy scales like $S\sim M^2 l_p^2$
and hence black holes dominate the spectrum.
Following \refs{\SusskindAA,\HorowitzNW}, let us consider the value
of mass for which the curvature near the horizon becomes order one in string
units, $M\sim l_s/l_p^2\sim 1/g_s^2l_s$.
The black hole entropy at this point, $S\sim M l_s$, is equal to that of perturbative
string, up to a numerical factor.

Recently D. Kutasov \KutasovRR\ proposed a way to determine the
correspondence point precisely.
Suppose we know the Schwarzschild solution within the string theory.
Knowing the solution to all orders in $\alpha'$ is important 
precisely because in the vicinity of the correspondence point  $\alpha'$ corrections
to the classical gravity become large near the horizon.
The euclidean solution is asymptotically flat and the circumference
of the temporal direction asymptotes to the inverse Hawking
temperature $\beta$.
According to \KutasovRR,  whenever $\beta$ is equal
to the Hagedorn temperature of perturbative strings, the entropies
of the black hole and fundamental strings agree.

To test these ideas \KutasovRR, one can look at the two-dimensional black hole,
which is described in string theory by an exactly solvable CFT.
The cleanest example is $\NN=2$ supersymmetric $SL(2)_k/U(1)$ with
central charge $c=3+6/k$, which should be dressed with an additional 
matter CFT to make string theory critical.
The black hole has a Hagedorn density of states (to be reviewed below)
with the Hagedorn temperature equal to the Hawking temperature,
\eqn\betabh{ {1\over T}=\beta=2\pi \sqrt{k \alpha'}  }
This should be compared with the Hagedorn temperature of perturbative
strings 
\eqn\betah{\beta_H=2\pi \sqrt{(2-1/k)\alpha'}  }
Hence, for $k>1$ the black holes dominate the spectrum, while at the
correspondence point $k=1$ determined by $\beta=\beta_H$, the entropies
of black holes and strings are, of course, the same \KutasovRR\foot{
Similar picture holds for charged two-dimensional black holes
\GiveonJV.}.
For $k<1$ black hole drops out of the physical 
spectrum  \GiveonMI.

In this paper we are mostly concerned with the theory at the correspondence point ($k=1$).
Consider the background
\eqn\bkgd{ \IR_t\times \IR_\phi \times \IS^1 \times \X}
where $\IR_t$ stands for the time direction,
$\IR_\phi$ is the linear dilaton theory with central charge
$c=1+3 Q^2/2$, and $\X$ is a compact SCFT with central charge $\hat c=3$.
Here and in the rest of  the paper we set $\alpha'=1$, hence
$Q^2=4/k=4$.
In \bkgd\  $\IS^1$ is a boson compactified at the 
self-dual radius, which is necessary for space-time supersymmetry \KutasovUA\
(for a recent discussion of non-critical superstrings see \ItzhakiZR).
To cap the strong coupling region of the linear dilaton theory we
introduce the $\NN=2$ Liouville perturbation in $\IR_\phi \times \IS^1$.
Equivalently, we modify the geometry \bkgd\ to
\eqn\bkgdm{ \IR_t\times {SL(2)\over U(1)} \times \X}
which asymptotes to \bkgd\ as $\phi\ra\infty$,
and set the string coupling at the tip of the cigar to a small but finite
value $g_s$.
The high energy density of perturbative string states in the background \bkgdm\
is given by
\eqn\hds{  \rho(M)\sim {e^{\beta_H M}\over M^{2}}  }
with $\beta_H$ given by \betah.
The appearance of $M^{2}$ in the denominator is due to the fact that 
the linear dilaton direction has a continuous spectrum of excitations above 
the gap.
Various derivations of \hds\ are reviewed in Appendix.

One can consider the euclidean black hole in the background \bkgdm.
When the black hole energy is large,
\eqn\bhel{ E>>g_s^{-2}   }
the geometry is well described by  
\eqn\bh{  {SL(2)_{k=1}\over U(1)}\times \IS^1\times \X    }
This is because the energy of the black hole is related to the coupling
at the tip of the thermal cigar, $SL(2)_{k=1}/U(1)$, as $E\sim \exp(-2\Phi_0)$,
and the effects of the original Liouville wall disappear in the limit \bhel.
In other words, the black hole geometry in \bh\ cuts off the linear dilaton direction 
way before the difference between \bkgd\ and \bkgdm\ becomes significant.
In the next section we review thermodynamics of \bh, which
exhibits Hagedorn density of states 
\eqn\hdsbh{  \rho_{BH}(E)\sim E^\alpha e^{\beta E}  }
with $\beta$ given by \betabh\ and $\alpha$ being a negative number.
In our case $k=1$ and hence $\beta=\beta_H$.

For energies sufficiently small so that both string self-interaction 
and non-perturbative effects are negligible, the perturbative closed string states are
the correct degrees of freedom, and the density of states is given by \hds.
Once the energy is raised to satisfy \bhel, the black hole picture becomes more
appropriate, and the density of states is described by \hdsbh.
Thus it is natural to expect that the entropy interpolates between the expressions
\hds\ and \hdsbh\ as the energy is varied.\foot{ In the BPS case the density of 
the perturbative states can be matched to that of the black hole beyond the leading order
in energy (for a recent discussion see \OoguriZV), due to non-renormalization.}
The situation is similar to the  higher dimensional case, where the description also interpolates 
between the perturbative string states and black holes as the energy is increased
\refs{\HorowitzNW,\DamourAW}.
The main feature of the two-dimensional case that we consider is that the
interpolation only happens in the subleading term,
since $\beta$ in \hdsbh\ is equal to $\beta_H$ in \hds.

It is interesting to compare $\alpha$ in \hdsbh\ with $-2$ 
which appears in \hds.
In the next Section we consider the simplest case of superstrings in
the background \bh\ with $\X=\IT^3$.
Hagedorn density of states \hdsbh\ implies that the free energy $\F$ of 
\bh\ is zero at leading order.
The first non-vanishing term is
\eqn\fe{  \beta \F=-(\alpha+1) \log E=-Z^{(1)}  }
where $Z^{(1)}$ is the one-loop string
partition function in the background \bh.
We compute  $Z^{(1)}$ and determine the value of $\alpha$.
When the volume of $\IT^3$ is large in string units, $|\alpha| >>1$.
The minimal value of $|\alpha|$ is achieved when all the compactification
radii take the self-dual values.
We discuss our results in Section 3.
Appendix contains a brief review of various ways
of computing and interpreting the string partition function.

%%%%%%%%%%%%%%%%%%%%%%%%%%%%%%%%%%%%%%%%%%%%%%%%%%%%%%%%%%%%%%%%%%%%%%%%%%%%%%%%%%
\newsec{String partition function in the black hole background}
In this section we compute the string partition function.
The discussion will closely follow \KutasovJP\ where analogous
computations were performed for Little String Theory, in
the regime $\beta_H<<\beta$.
We are interested in the  $\beta_H=\beta$ case, where 
$Z^{(1)}$ involves a complicated integral.
Although one can correctly estimate the behavior of the integrand,
the value of the integral can only be obtained numerically.

The bosonic matter content of the theory is given by \bh\ with $\X=\IT^3$.
The energy of the black hole is related to the coupling
at the tip of the cigar as $E\sim \exp(-2\Phi_0)$, with the precise
coefficient being unimportant.
Far from the tip of the cigar, the $SL(2)_1/U(1)$ is described
by a product of a thermal circle and a linear dilaton,
$\IS^1_\beta\times \IR_\phi$.
In addition, there are six free fermions; thermal boundary
conditions along the $\IS^1_\beta$ are implemented in
the usual way \AtickSI.
The theory enjoys $\NN=2$ worldsheet supersymmetry, which 
ensures vanishing of the genus zero partition function \KutasovJP.
This is in accord with Hagedorn density of states \hdsbh:
at leading order the temperature is independent of the energy and
$\beta\F=\beta E-S=0$.
The one-loop string partition function is in general non-zero, and
proportional to the volume of the linear dilaton direction 
(which, in turn, is proportional to $\log E$).
The computation, and the resulting thermodynamic behavior, is similar
to the Little String Theory case which has been studied in 
\refs{\KutasovJP,\MaldacenaCG\AharonyTT\HarmarkHW\BerkoozMZ\RangamaniIR\BuchelDG\NarayanDR\DeBoerDD-\AharonyXN}.

The spacetime supersymmetric one-loop partition function at zero temperature can be found in
\MizoguchiKK:
\eqn\pfzerot{  Z= V_{\IR_t\times\IR_\phi} \int_{\F_0} {d^2\tau\over2\tau_2} 
                 {1\over (4\pi^2 \tau_2)} {1\over4|\eta(\tau)|^{12}}\left(
                 |\Lambda_1(\tau)|^2+|\Lambda_2(\tau)|^2  \right) \prod_{i=1}^3 \Theta(R_i,\tau) }
where
\eqn\lamone{\Lambda_1(\tau)=\Theta_{1,1}(\tau)\left(\theta_3^2(\tau)+\theta_4^2(\tau)\right)
                            -\Theta_{0,1}(\tau)\theta_2^2(\tau)   } 
and $\Lambda_1(\tau)$ is related to its modular transform under $\tau\ra -1/\tau$,
\eqn\lamtwo{\Lambda_2(\tau)=\Theta_{0,1}(\tau)\left(\theta_3^2(\tau)-\theta_4^2(\tau)\right)
                            -\Theta_{1,1}(\tau)\theta_2^2(\tau)   } 
In \pfzerot\ --\lamtwo\
\eqn\thetabig{ \Theta_{m,1}=\sum_{n\in \IZ} e^{2\pi i \tau (n+m/2)^2}         }   
and 
\eqn\thetai{  \Theta(R_i,\tau)=\sum_{n,w=-\infty}^\infty
               \exp\left[ -\pi\tau_2({n^2\over R^2}+w^2 R^2)+2\pi i \tau_1 n w \right]    }
Partition function \pfzerot\ corresponds to the GSO projection by $(-)^{F+F_{st}}$
where $F$ and $F_{st}$ are the worldsheet and spacetime fermion numbers, respectively.
There are four massless bosons in both the NSNS and RR sector.
Both \lamone\ and \lamtwo\ are equal to zero, so that the partition
function \pfzerot\ vanishes as expected from spacetime supersymmetry.
 
Finite temperature results in the extra summation over momentum
and winding with respect to the thermal circle.
Let us introduce
\eqn\lamonenm{\Lambda_1^{(n,m)}(\tau)=\Theta_{1,1}(\tau)    
     \left[U_3(n,m) \theta_3^2(\tau)+U_4(n,m)\theta_4^2(\tau)\right]
                            -\Theta_{0,1}(\tau)U_2(n,m)\theta_2^2(\tau)   }
and
\eqn\lamtwonm{\Lambda_2^{(n,m)}(\tau)=\Theta_{0,1}(\tau)  
     \left[U_3(n,m)\theta_3^2(\tau)-U_4(n,m)\theta_4^2(\tau)\right]
                            -\Theta_{1,1}(\tau)U_2(n,m)\theta_2^2(\tau)   } 
where $U_\mu(n,m)$ are the phase factors which can be found in \AtickSI:
\eqn\uphase{
\eqalign{
  U_1(n,m)&={1\over2}\left(-1+(-1)^n+(-1)^m+(-1)^{n+m}\right) \cr
  U_2(n,m)&={1\over2}\left(1-(-1)^n+(-1)^m+(-1)^{n+m}\right)  \cr
  U_3(n,m)&={1\over2}\left(1+(-1)^n+(-1)^m-(-1)^{n+m}\right)  \cr
  U_4(n,m)&={1\over2}\left(1+(-1)^n-(-1)^m+(-1)^{n+m}\right)  \cr}
}
The free energy \fe\ is given by
\eqn\pft{  \beta \F={-} \beta L_{\phi} \int_{\F_0} {d^2\tau\over 32 \pi^2 \tau_2^2|\eta(\tau)|^{12}}
              \sum_{m,n=-\infty}^\infty \left(
                 |\Lambda_1^{(n,m)}(\tau)|^2{+}|\Lambda_2^{(n,m)}(\tau)|^2  \right) 
                 e^{-S_\beta(n,m)}
                 \prod_{i=1}^3 \Theta(R_i,\tau) }
where the volume of the linear dilaton direction 
\eqn\lphi{ L_\phi=-{1\over Q}\log E+{\rm const} \cong -{1\over 2} \log E   }
and
\eqn\sbeta{  S_\beta(n,m)=-{\beta^2|m-n\tau|^2\over 4\pi\tau_2}   }
Using modular invariance of the integrand in \pft,
the sum over $n$ can be restricted to $n=0$, and the integration 
domain changed to $-1/2<\tau_1<1/2,\;0<\tau_2<\infty$.
The sum over even $m$ in \pft\ vanishes.
Substituting $\beta=2\pi$ in \pft\ and comparing with $\fe$ we have
\eqn\alphaone{ \alpha+1=-\int {d^2\tau\over 32 \pi \tau_2^2|\eta(\tau)|^{12}}
              \sum_{m\in 2\IZ+1} \left(
                 |\Lambda_1^{(0,1)}(\tau)|^2{+}|\Lambda_2^{(0,1)}(\tau)|^2  \right) 
                 e^{-{\pi m^2\over\tau_2}}
                 \prod_{i=1}^3 \Theta(R_i,\tau) }
To estimate the behavior of the integrand as $\tau_2\ra\infty$
note that only $n=w=0$ terms in \thetai\ contribute in this regime.
The sum over $m$ in $\alphaone$ needs to be Poisson resummed, giving
an extra factor of $\sqrt{\tau_2}$.
The terms corresponding to the massive states are exponentially suppressed.
To summarize, as $\tau_2\ra\infty$ the integrand in \alphaone\ is dominated
by massless states and goes as $\tau_2^{-3/2}$.
In the case studied in \KutasovJP\ this was the end of story, as
$\beta$ was much larger then $\beta_H$ and hence the integrand was
heavily suppressed near $\tau_2=0$.
It is no longer the case, and the integrand, in fact,  behaves like $\tau_2^{-1/2}$
as $\tau_2\ra0$.
This is in accord with the behavior of the string partition function
as $\beta\ra \beta_H$, as explained in Appendix.
One can also see this directly from \alphaone.
The naive power counting of $\tau_2$ gives
\eqn\taupower{  \tau_2^{-1} \tau_2^{-1} \tau_2^{-4} \tau_2^4\sim \tau_2^{-2}   }
where the first factor comes from the measure, the second from 
the $\IR_\phi\times\IS^1_\beta$, the third from the $\IS\times\IT^3$, and the last
from the combinations of $\theta_\mu(\tau)$ and $\eta(\tau)$.
One must be more careful, however, as there is an important exponential
\eqn\impexp{  \exp\left({\pi i\over\tau}-{\pi\over\tau_2}\right)\sim 
          \exp\left(-{\pi\tau_1^2\over |\tau|^2 \tau_2}\right) }  
which stays constant along the curves $\tau_1=x \tau_2^{3/2}$ parameterized by $x$.
Switching to the variables $x$ and $\tau_2$ one earns the Jacobian 
$J=\tau_2^{3/2}$.
Multiplying this by \taupower\ gives the stated behavior near $\tau_2\sim 0$.

All string states contribute to $\alpha$ in \alphaone\ and no further
simplifications occur.
Generally the right-hand side of \alphaone\ is proportional to the
compactification volume, therefore at large volume $\alpha$ is bound to
be a large negative number.
The minimum of $|\alpha|$ is achieved when all the radii take the self-dual values, $R_i=1$.
The integral at this point can be evaluated numerically, giving $\alpha=-1.68$.
%We can also determine the values of $R_i$ when $\alpha=-2$, and the subleading
%term in the entropy has the same asymptotics at small and large energies.
%This happens when $R_i^2\approx3.0$.

\newsec{Discussion}

In this paper we considered a particular case of spacetime-supersymmetric
noncritical strings in the background \bkgdm\ whose perturbative Hagedorn temperature
is equal to the Hawking temperature of the black hole which asymptotes to
\bkgdm\ at infinity.
This required choosing the matter content with $\hat c=4$ to dress the linear dilaton
theory with $Q=2$.
To cut off the strong coupling region we introduced the $\NN=2$ Liouville wall in 
the background \bkgd\ or, equivalently, considered the geometry \bkgdm\ with small but finite coupling
$g_s$ at the tip of the $SL(2)/U(1)$ cigar.
Massive black holes in this background are described by the euclidean geometry \bh.
The subleading correction to the black hole entropy is related to
the value of the string one-loop partition function in the background \bh
\foot{
The perturbative calculation of the black hole entropy is, strictly
speaking, only valid for temperatures larger than $\beta_H^{-1}$.
Nevertheless, the microcanonical description is well defined 
for both strings and black hole. 
Using formal inverse Laplace transform to obtain the value of $\alpha$
might be a point of concern.
We have nothing new to say regarding this issue.}.

In the black hole/string transition picture 
of Refs. \refs{\SusskindAA,\HorowitzNW,\DamourAW} the correct degrees of
freedom interpolate between the perturbative string states 
and black holes as the energy is increased.
More precisely, the gravitational self-interaction of the string
becomes important when $M>1/g_s^{4\over6-d}$ where $d$ is the number
of non-compact spatial dimensions \HorowitzNW.
In addition, non-perturbative states become important when $M>1/g_s$.
In our case $d=1$, so both effects make the perturbative string density
of states \hds\ unreliable way before the black hole becomes
a good description at $M>>1/g_s^2$.
We expect the density of states to interpolate between the perturbative
string expression [Eq. \hds] and its black hole counterpart [Eq. \hdsbh]
as the energy is increased.
Since the leading exponential behavior is the same, thanks to $\beta=\beta_H$,
the interpolation apparently takes place in the subleading term.
This should be contrasted with the higher-dimensional case, where
the leading behavior of the entropy is different for perturbative strings 
and black holes.

As a specific model, we considered superstrings in the background
\bkgdm\ with $\X=\IT^3$.
The value of $\alpha$ depends on the volume of compactification.
Compactification at the self-dual radii yields $\alpha=-1.68$.
When the volume of $\IT^3$ becomes large,
$\alpha$ scales like $ -V_{\IT^3}$\foot{ $\alpha=-2$ when $R_i^2\approx 3.0$.}.
Other choices of $\hat c=4$ matter may lead to other 
values of $\alpha$. 
In addition to superstrings, we have also considered type 0
and bosonic strings compactified on $\IT^4$ and $\IT^{18}$ respectively,
where similar black hole/string transitions occur.
These cases, however, are pathological due to the presence of tachyon
in the spectrum, which has to be removed by hand.

It is interesting that the numerical value of $\alpha$ at
the self-dual compactification radii computed in the previous Section ($\alpha=-1.68$)
is not very different from the corresponding
quantity for the perturbative strings ($\alpha=-2$), and therefore even the subleading term 
in the entropy does not change drastically as the energy is varied.

What happens when $k>1$ and the leading terms in the black hole and string
entropies no longer agree? As before, we cut off the strong coupling
region by introducing a cigar geometry with the string coupling at the
tip of the cigar equal to a small but finite value $g_s$.
The picture is similar to the $k=1$ case discussed above.
When $M>>1/g_s^2$, the black holes are good degrees of freedom
and the Hagedorn temperature is now smaller than that of perturbative strings.
For sufficiently small energies 
the perturbative strings become the correct degrees of freedom.
We expect the entropy to interpolate between the perturbative string and
the black hole expressions.

\bigskip\bigskip\noindent{\bf Acknowledgements:}
We thank T.~Banks, G.~Moore and especially D.~Kutasov for useful discussions. We also thank the organizers of the
``Mathematical Structures in String Theory'' workshop and KITP, UC Santa Barbara
for hospitality while this work has been completed.
This work was supported in part by  DOE grant DE-FG02-96ER40949 for Rutgers 
and NSF grant PHY99-07949 for KITP.

%%%%%%%%%%%%%%%%%%%%%%%%%%%%%%%%%%%%%%%%%%%%%%%%%%%%%%%%%%%%%
\bigskip\bigskip\appendix{A}{Comments on one-loop string partition function}

Standard arguments \GreenSP\ imply that the number of closed 
string states at level $n$ in superstring theory compactified to
$D$ spacetime dimensions is (see e.g. \DienesHX)
\eqn\dstates{  d_n\sim n^{-{1\over2}(D+1)} e^{C\sqrt{n}}   }
where $C$ is an easily computable constant which determines the Hagedorn temperature of perturbative strings
in the linear dilaton background \betah.
We are interested in the $D=2$ case.
Eq. \dstates\ translates into the mass density $\rho(M)\sim M^{-2} e^{\beta_H  M}$
The contribution of high energy states to the free energy of the string  
gas is
\eqn\fesg{  \log Z\sim \int dM M^{-2} \exp\left(\beta_H M\right)\int_{0}^\infty dk\;
                               \exp\left(-\beta\sqrt{k^2+M^2}\right)    }
We are interested in the behavior of \fesg\ as $\beta\ra\beta_H$:
\eqn\fesga{  \log Z\sim \int dE E^{-3/2} \exp\left( (\beta_H-\beta) E\right)\sim (\beta-\beta_H)^{1/2}  }
The same answer follows from the thermal scalar calculation \HorowitzNW,
where the power of $(\beta-\beta_H)$ is determined by a number of 
dimensions with continuos spectra of excitations.

According to Polchinski \PolchinskiZF\ \fesg\ is equal to the one-loop 
string partition function in the background \bkgd.
This is what we were computing in Section 2, and indeed the behavior 
\fesga\ is consistent with the $\tau_2\ra 0$ asymptotics of the 
integrand in \alphaone:
\eqn\consi{   \int d\tau_2 \tau_2^{-1/2}  
       \exp\left( {\beta_H-\beta\over\tau_2}\right)\sim (\beta-\beta_H)^{1/2} } 
The fact that \fesga\ vanishes as $\beta\ra\beta_H$ means that all states 
contribute to the free energy giving a constant which has been
omitted in \fesga; that is why we had to resort to
numerical methods to evaluate the integral \alphaone.

%%%%%%%%%%%%%%%%%%
\listrefs

\end